\documentclass{PoS}

\usepackage{xspace}
\usepackage{amsmath}
\usepackage{url}

\newcommand{\DO}{{D{\O}}\xspace}

\newcommand{\eVdist}{\kern-0.06667em}
\newcommand{\Mev}{{\text{Me}\eVdist\text{V\/}}}
\newcommand{\Gev}{{\text{Ge}\eVdist\text{V\/}}}
\newcommand{\Tev}{{\text{Te}\eVdist\text{V\/}}}

\newcommand{\pb}{\,\text{pb}}
\newcommand{\fb}{\,\text{fb}}

\newcommand{\pbi}{\,\text{pb}^{-1}}
\newcommand{\fbi}{\,\text{fb}^{-1}}

\title{Summary of the ``Electroweak and Searches in DIS and Hadron Colliders'' Working Group}

\ShortTitle{Electroweak and Searches: Working Group Summary}

\author{Stefano Moretti\thanks{This work is supported in part by the NExT Institute and SEPnet.}\\
        School of Physics \& Astronomy, University of Southampton\\
        Highfield, Southampton SO17 1BJ, United Kingdom\\
        E-mail: \email{S.Moretti@soton.ac.uk}
}
\author{Andrea Parenti\\
        Deutsches Elektronen-Synchrotron\\
        Notkestra{\ss}e 85, 22607 Hamburg, Germany\\
        E-mail: \email{parenti@mail.desy.de}
}

\abstract{We present here the summary of the ``Electroweak and Searches in DIS and Hadron Colliders'' working group activities at the DIS 2010 conference. The theory contributions will be presented first, followed by the experimental ones. The experimental results include measurements from the Tevatron (CDF and \DO) and HERA (H1 and ZEUS) experiments, and studies from the LHC (ATLAS and CMS). Results from the B-factories (BaBar and Belle) and from NA62 are also summarised.}

\FullConference{XVIII International Workshop on Deep-Inelastic Scattering and Related Subjects, DIS 2010\\
		April 19-23, 2010\\
		Firenze, Italy}

\begin{document}

\section{Introduction}
Despite the eruption of the Eyjafjallaj\"okull volcano in Iceland, whose ashes caused the complete stop of all flights in northern Europe in the days preceding the conference, the electroweak and searches session was very successful. We had 43 presentations, some of which in combined sessions with other Working Groups (WGs): six together with the parton densities WG, five with the QCD final states WG and four with the future of DIS WG. Some of the above mentioned talks were given in video conference, but only seven had to be cancelled.

\section{Theoretical Summary}
A new era in particle physics is beginning, spurred on by the LHC starting its data taking phase at CERN. Building on the success of current DIS and hadron colliders (e.g., HERA at DESY, Tevatron at FNAL), the physics expectations at the new hadron machine are enormous. In particular, from the point of view of theoretical particle physics, it will be a defining moment, as the energy ranges afforded by the LHC (from 7 to $14~\Tev$) would ultimately enable one to confirm or disprove the Standard Model (SM) as a viable scenario for the $\Tev$ scale. This would in fact require the existence of a rather light Higgs boson. Failing this, the requirement of perturbative unitarity of whichever underlying theory would impose the existence of some form of new physics Beyond the SM (BSM). Hence, unsurprisingly, all of the theory talks presented in this working group tackled one or more of these aspects. 

The plan of this summary is then to review these talks, by emphasising the common guiding threads existing amongst all of them. We will do so in the next section, where we will highlight the most important results of each contribution. Finally, based on the scope of the latter, we will conclude by providing a common outlook to all of the perspectives presented.   

\subsection{Higgs or non-Higgs, this is the question} 
The key question addressed essentially by all of the talks presented in the working group, either directly or indirectly, is whether Electro-Weak Symmetry Breaking (EWSB) is based on a Higgs-type mechanism (hence within a weakly coupled framework) or else on alternative (Higgsless) dynamics (typically realised in stron\-gly coupled theories) and in either case to understand which actual realisation takes place in nature.

In the attempt to secure the possibility of the LHC of accessing, if existing, a Higgs boson signal in all viable channels, Ref.~\cite{Zaro} presented next-to-next-to-leading order (NNLO) QCD corrections to Higgs production via the Vector Boson Fusion (VBF) channel, which can in fact be used both as a discovery mode of and a means of profiling a Higgs boson through several of its subsequent decay channels. Fig. 1 makes the point that a clear convergence of the perturbative series is achieved at LHC energies of $14~\Tev$ (similar results occur at $7~\Tev$), as $\Delta {\rm NNLO}/{\rm NLO} = {\cal O}(1\%)$ and the residual theoretical uncertainities, i.e., the renormalisation/factorisation scale dependence and that upon the Parton Distribution Functions (PDFs), reduce to $1-2\%$ in the NNLO cross section.

In a rather similar vein, Ref.~\cite{Santos} dealt instead with the hitherto neglected possibility of extracting a light Higgs boson signal at the LHC using a novel detection channel based on a $\tau$ signature, which could indeed be assigned to the SM or else to some extensions of it based on an extended Higgs sector. Specifically, the process $pp (gg, gq ) \to h ~+~{jet}\to  \tau^+ \tau^-~+~ jet$ was used, by exploiting both leptonic and semi-leptonic $\tau$ decays, and a thorough signal-to-background analysis was carried out, showing that a SM Higgs resonance in the region of $120~\Gev$ can be extracted at the LHC, see Fig. 2 (e.g., for the leptonic case, though similar results can be obtained in the case of semi-leptonic decays too). In fact, in the case of scenarios with an extended Higgs sector, like pure 2-Higgs Doublet Models (2HDMs) or even the ones obtained by adding an arbitrary number $n$ of doublets that {\sl do not couple} to fermions (labelled as 2HDM+$n$D), a signal can be established also for much smaller Higgs masses, down to $50-60~\Gev$. Similar results also hold  in the case of the so-called democratic three-Higgs doublet model (acronymed as 3HDM(D)) where up-type quarks, down-type quarks and charged leptons all get their mass from a different doublet.  

Amongst models with extended Higgs sector, over the years, there has been a prevalence of Supersymmetry (SUSY) based realisations, like the Minimal Supersymmetric Standard Model (MSSM), which -- from the point of view of the Higgs sector at tree level -- is a nothing but a particular version of a 2HDM, a so-called type II, with specific relations between Higgs masses and couplings enforced by SUSY. The MSSM though, despite incorporating all of the benefits of SUSY (gauge coupling unification, a solution to the hierarchy problem, a dark matter candidate, etc.) and, thanks to its minimal particle content, being very predictive (particularly in its definite theoretical upper bound on the lightest Higgs boson mass, of about $130~\Gev$), suffers from the problem that much of its parameter space is ruled out by experiment because of the failure of LEP (and Tevatron) in finding such a light object in direct searches. While several non-minimal realisations of SUSY, which could alleviate this problem, have been proposed to date, where the assumption of `minimality' is dismissed for the gauge group, the particle content and/or the Higgs representation, none of these has a profound theoretical basis and certainly is not supported by experiment. Hence, to develop instead an effective field-theory approach, assuming that the MSSM is valid up to a heavy physics scale $M$ (the so-called Beyond the MSSM, or BMSSM, approach), which is essentially model independent and, in particular, parametrises deviations from the MSSM up to terms of order $1/M^2$, seems as an effective method to guide us in eventually disentangling the true realisation in nature of SUSY. This is what has been discussed in Ref.~\cite{Zurita}. In BMSSM Higgs sectors, the collider phenomenology can be greatly different from that of both the SM and the MSSM. In particular, with respect to the latter,  the most striking feature is that a sizable rise of the lightest CP-even Higgs mass can be attained (especially for low $\tan\beta$'s), thus relaxing the aforementioned tension in this respect in the MSSM context, see Fig. 3. Besides, in the large $\tan\beta$ regime the BR($h\to\gamma\gamma$) can sizably be enhanced in the BMSSM relative to the SM (and consequently the MSSM), an intriguing result in view of searches for light Higgs bosons at both the Tevatron and the LHC.

If one dismisses the Higgs hypothesis, and then ventures into some Higgsless realisation of EWSB, the possibilities are several and these have been reviewed at this conference in Ref.~\cite{Grojean}\footnote{The case of composite Higgs solutions was also illustrated here as another viable alternative realisation of EWSB.}. In the most general (Higgsless) framework of a strongly interacting dynamics for EWSB, one typically has non-fundamental (i.e., composite) resonances (needed, if anything, in keeping under control the longitudinal SM gauge boson scattering up to some $\Tev$ scale cutoff, yet complaint with EW precision tests) of various spin (0, 1/2 and 1), held together by a new strong interaction. Ref.~\cite{Carcamo} specifically dealt with the case of a triplet of composite vectors belonging to the adjoint representation of a custodial symmetry group $SU(2)_{L+R}$, which could be produced in pairs at the LHC via both the Drell-Yan (DY) process and VBF. Fig. 4 illustrates the latter case at the LHC for $\sqrt s=14~\Tev$. Results for the DY channel are similarly interesting, although somewhat smaller in total rates. The emerging signatures for these new particles would involve di- and tri-lepton final states, with and without accompanying jets, which may in principle be extracted from SM background, for composite vector $V$ masses $M_V$ of order $500~\Gev$ or so, although a detailed phenomenological analysis in this respect is still lacking.

While new physics may be needed at the $\Tev$ scale, whether due to some enlarged Higgs sector or alternative EWSB mechanism, this could well not manifest itself in resonant production of new particle states, for example, when their mass is beyond the kinematic reach of the collider or their couplings (to the ordinary matter present in the (anti)protons) are suppressed. Under these circumstances, it becomes of paramount importance to assess the scope of `precision physics' at hadronic machines, in a similar spirit to the time honoured one adopted at LEP, wherein the existence of top quarks could not be confirmed directly by data (owning to the large value of the top mass with respect to the collider energy, so that a top quark could not be resonantly produced) yet it could be inferred indirectly and rather accurately by precision measurements of EW observables. In this connection, some remarks are in order though. The Tevatron and the LHC are not commonly perceived as precision machines, due to both the undefined partonic energy and the intrinsically large hadronic background, so that virtual effects of new physics are expected to be not easily discernible in the data. However, there are certain processes where the experimental precision is expected to eventually become comparable to the size of the possible virtual effects of new physics. Among these processes, one can certainly list top quark production, both in double- and single-top mode, owning to a (ultimately) small experimental error, both statistical (because of the large top production cross sections and luminosities to be attained, particularly at the LHC)  and systematic (dominated by the jet energy scale uncertainty, which is being effectively overcome at Tevatron and will be at LHC too after the first hadronic data samples will have been collected and studied in detail).

Therefore, it is mandatory to keep the theoretical error under control as well (this is intrinsically due to missing higher order QCD corrections, affecting both the PDFs and the hard scattering matrix element) in the ultimate attempt to understand the underlying dynamics of EWSB. Refs.~\cite{Ferrario} and \cite{Kidonakis} contributed to this effort. The former paper dealt with the case of double-top production, by building on previous results from NLO QCD which estimated the size of such effects onto the charge asymmetry of top quarks, it showed that possible new EWSB physics effects which manifest themselves through heavy coloured resonances decaying to top-antitop quark pairs (due, e.g., to axigluons in chiral colour models, different kinds of massive gauge bosons in colouron and top colour models or Kaluza Klein excitations in models with extra spatial dimensions) could be extractable at both Tevatron (directly in $t\bar t$ samples) and the LHC (also in presence of additional gluon radiation in $t\bar t$ + $jet$ samples) from the study of such an asymmetry. Fig. 5 makes this point for the illustrative example of a model independent heavy colour-octet resonance $G$ of mass $m_G=1.5~\Tev$ and generic vector and axial couplings at the LHC. Prospects in this respect are good at both a 7 and $14~\Tev$ LHC (with expected statistics) while at Tevatron a much increased luminosity with respect to the present would presumably be required. The latter paper, based on recent developments in NNLO calculations of soft anomalous dimensions, performed a resummation through next-to-next-to-leading-logarithm (NNLL) accuracy of QCD effects onto the `$s$-channel' single-top production in $t\bar b$ final states as well as onto the `associated production' of a single-top with a $W^\pm$\footnote{In fact, also the associated production of a single-top with a charged Higgs boson was tackled here.}, at both the Tevatron and the LHC, confirming good theoretical control of the predictions, in both the inclusive and exclusive cross section. $K$ factors through (approximate) NNLO are reasonably small, increasing by some 10\% the (exact) NLO ones at both Tevatron and LHC for $t\bar b$ production (Fig. 6 shows the LHC case for $14~\Tev$). In the case of $tW^-$ production the approximate NNLO corrections increase the exact NLO results by about 8\% at both colliders (in the $tH^-$ case they are twice as big).

\subsection{Outlook}
EWSB requires Goldstone boson-type particles to provide mass to the $W^\pm$ and $Z$ gauge bosons and at the same time EW interactions taking place at the $\Tev$ scale need a light scalar state or some other (BSM) physics (essentially in the form of an ultraviolet regulator) to unitarise gauge boson scattering amplitudes. If not already the upgraded Tevatron with increased luminosity (as recently announced), certainly the LHC with standard energy and luminosity designs, will provide a solution to the puzzle of EWSB. This may appear in the form of a fundamental light Higgs boson, either of SM or BSM nature or alternatively be of Higgsless nature (composite (Higgs and vector) resonant states could also provide an alternative solution). The present working group has therefore occupied itself, from a theoretical and phenomenological perspective, with several topics related to EWSB, including the following two situations: where the solution to EWSB could be manifest in direct production of old and new particle states or else appear virtually in SM-type processes (chiefly involving top quark production in both single- and double-top channels). The results presented in the DIS 2010 sessions of the  `Electroweak and Searches in DIS and Hadron Colliders' working group unmistakably point towards confirming that several well advanced theoretical scenarios of EWSB exist, which are testable at present and future hadronic machines in both standard searches and precision measurements, as to the former they provide sizable event rates in signatures which should be extractable from pure SM backgrounds and since the latter are experimentally feasible (at least in top quark samples) and theoretically motivated (since QCD corrections are now generally under control).

\begin{figure}[t!]
  \centering
  \includegraphics[height=\textheight]{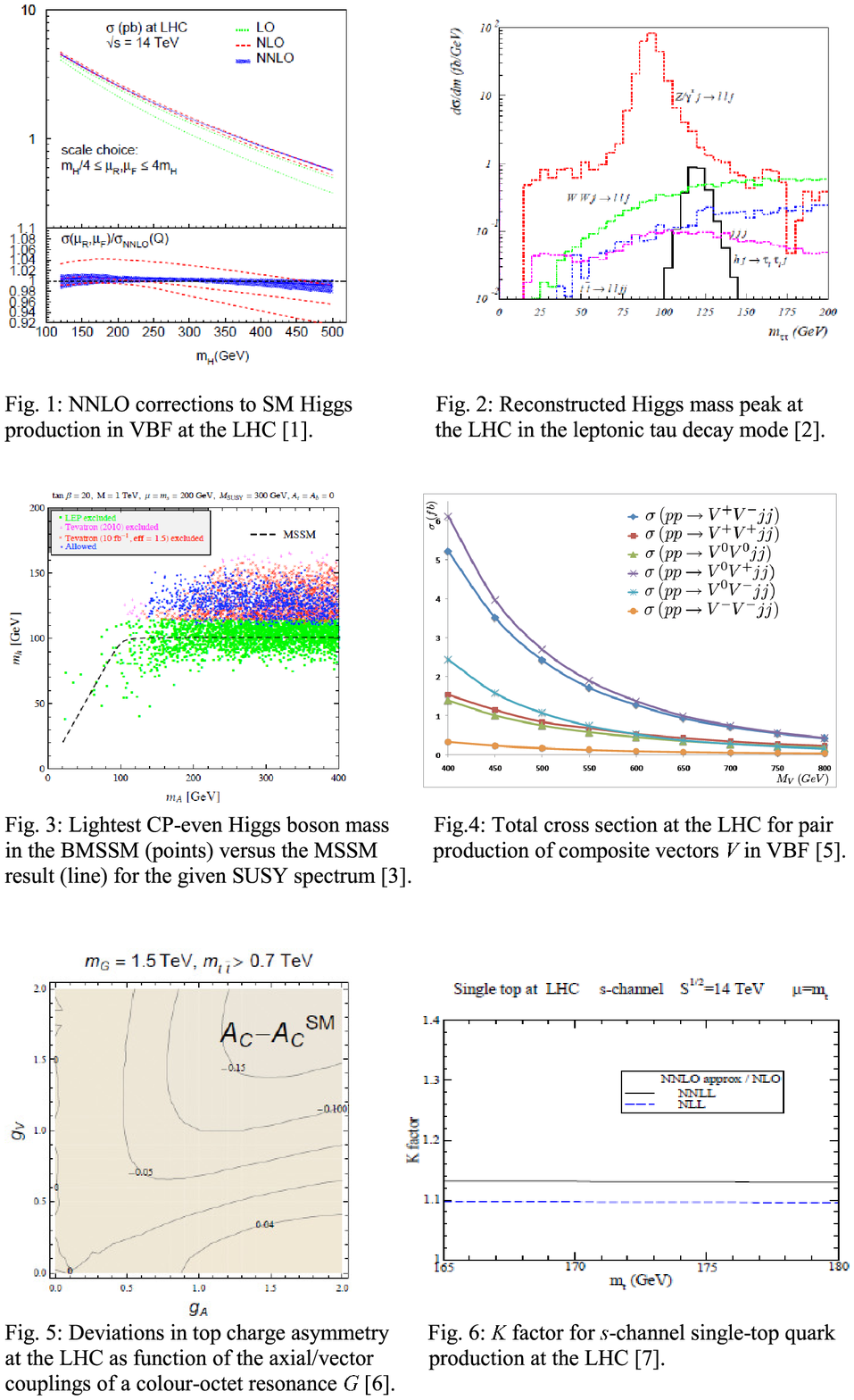}
\end{figure}
\addtocounter{figure}{6}

\section{Experimental Summary}

A large fraction of the experimental talks in the electroweak and searches session presented results from the Tevatron, LHC and HERA experiments. The Tevatron performances are impressive: as of July 2010 there were approximately $9~\fbi$ of $p \bar{p}$ collisions already delivered by the machine at a centre-of-mass energy $\sqrt{s} = 1.96~\Tev$, and the results shown by CDF and \DO during the conference were based on up to $\sim 5~\fbi$. On the other hand LHC is also coming into the game: it is colliding proton beams at $\sqrt{s} = 7~\Tev$ since March 30, and plans to collect $1~\fbi$ by the end of 2011. Finally, HERA was shut-down in summer 2007 and delivered approximately $0.5~\fbi$ of $e^\pm p$ collisions per experiment, at a centre-of-mass energy of $301-319~\Gev$; the first results based on the full statistics have already been published, and more will come in the next months and years.

In this section we will present an overview of the experimental results that were shown in the electroweak and searches session.

\subsection{$W$ and $Z$ production and properties}

\begin{figure}[!ht]
  \centering
  \includegraphics[width=0.30\textheight]{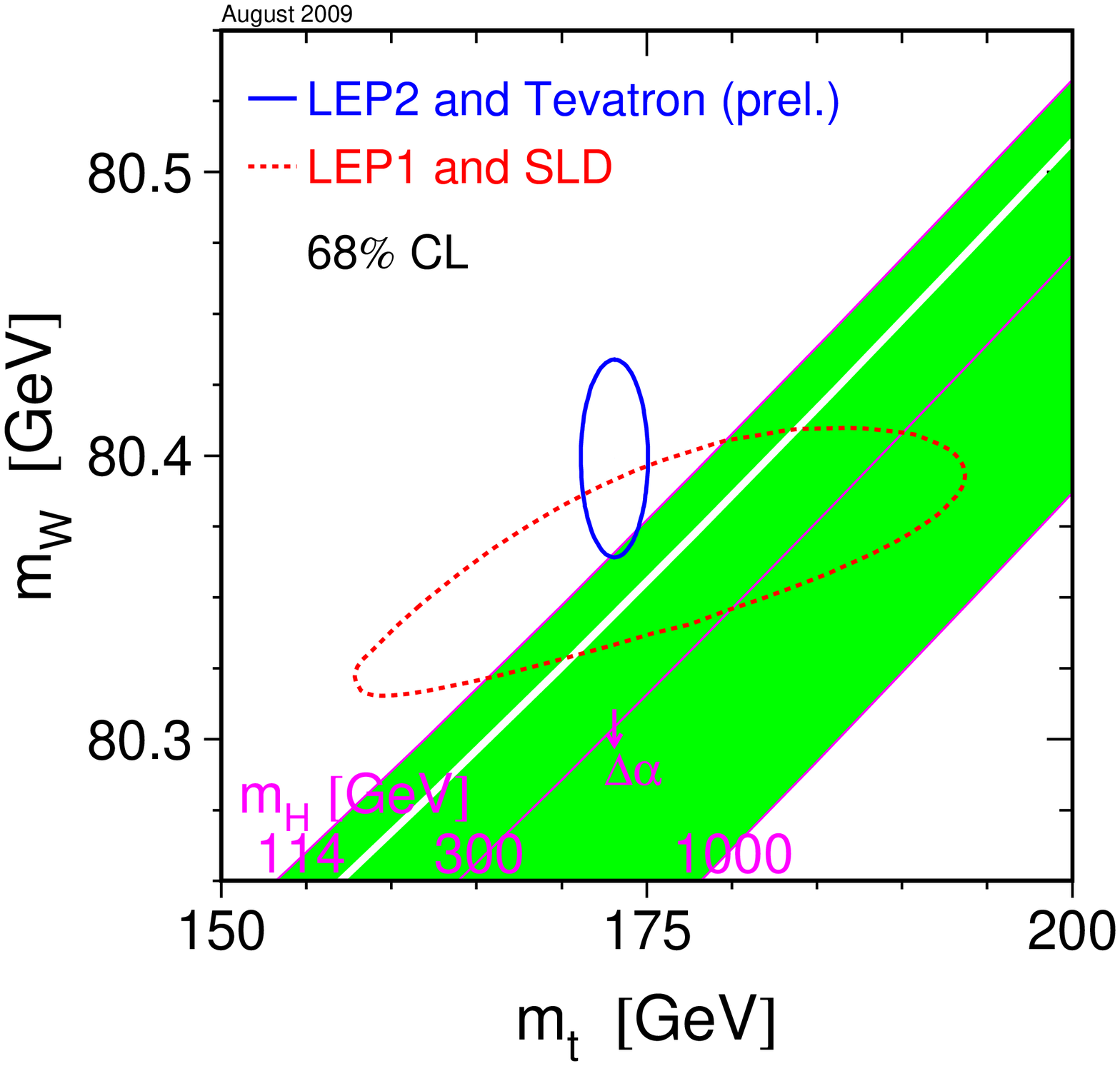}
  \caption{The comparison of the indirect constraints on $m_W$ and $m_t$ based on LEP-I/SLD data (dashed contour) and the direct measurements from the LEP-II/Tevatron experiments (solid contour). In both cases the 68\% C.L. contours are plotted. Also shown is the SM relationship for the masses as a function of the Higgs mass in the region favoured by theory ($< 1000~\Gev$) and allowed by direct searches ($114~\Gev$ to $170~\Gev$ and $> 180~\Gev$).
 The arrow labelled $\Delta \alpha$ shows the variation of this relation if $\alpha(m^2_Z)$ is changed by plus/minus one standard deviation. This variation gives an additional uncertainty to the SM band shown in the figure.}
  \label{fig:mtmw}
\end{figure}
The $W$ and $Z$ properties have been extensively studied at the Tevatron~\cite{chiarelli}, thanks to the large collected statistics. The very large samples of vector bosons -- more than 600000 leptonically decaying $Z$'s per experiment, and ten times more $W$'s -- also allows to accurately measure SM parameters. From the forward-backward asymmetry in $Z$ events, $A_{fb}$, it is possible to extract $\sin ^2 \theta_W$; \DO already determined $\sin ^2 \theta_W$ with a precision of 0.002 with $1~\fbi$ of data, and it is expected to reach the current world average precision, 0.0007, by combining the $8~\fbi$ already on tape for the two experiments. Moreover the charge asymmetry in $W$ decays, $A_W$, can help in constraining the proton structure functions, in particular the $d$ valence quark content at large $x$. Finally, the precision of the $m_W$ measurement is also of great importance since -- together with the top mass -- it constraints the SM Higgs mass, Figure \ref{fig:mtmw}. The total uncertainty on $m_W$ is $31~\Mev$ for the latest Tevatron combination, already better than the LEP-II average, which had an uncertainty of $33~\Mev$, and approaching that of the world average, $23~\Mev$.

At the LHC the $W$ and $Z$ bosons have a very high production cross section, and are the most important source of high-$p_T$ leptons. The vector boson production will be an important process in the first stages of the data taking; the $Z$ decaying into lepton pairs is suitable for the determination of trigger and reconstruction efficiencies, as well as for evaluating detector resolution and performing alignment. The processes can be also used for the estimation of the data luminosity. Finally, the charge asymmetry of the $W$ leptonic decays can be used to improve the determination of the proton structure functions at low-$x$.
It has been shown~\cite{borroni} that already with a data sample of $50~\pbi$ collected by ATLAS with a centre-of-mass energy of $14~\Tev$, a large sample of $W$ and $Z$ bosons will be collected. A broad range of measurements will be already feasible with $100~\pbi$ collected at a centre-of-mass energy of $7~\Tev$, that means by the end of 2010.

Methods for an early measurement of $W$ and $Z$ cross sections using $10~\pbi$ of proton-proton collision collected with the CMS detector at a centre-of-mass energy of $10~\Tev$ were presented~\cite{lenzi}. Monte Carlo events were selected by requiring one (in the case of $W$) or two (in the case of $Z$) high-$p_T$ leptons (electrons or muons) and large missing transverse energy (MET) or large transverse mass ($M_T$). The cross section times the branching ratio for $Z$ decaying into electrons could be determined with a statistical uncertainty better than 2\%, while the dominant systematic uncertainty will be due to the luminosity measurement, that will contribute with approximately a 10\%. Also for the $W$ production the systematic uncertainties will be dominated by the luminosity uncertainty.

The search for events with isolated leptons and missing transverse momentum, done at HERA by the H1 and ZEUS collaboration with $0.98~\fbi$ of $e^\pm p$ collisions, has been published. 81 events have been observed, while $87.8 \pm 11.0$ were expected from SM processes, dominated by single $W$ production. The single $W$ production cross section was measured to be  $\sigma = 1.06 \pm 0.16 (\mathrm{stat.}) \pm 0.07(\mathrm{syst.})~\pb$, in agreement with the SM prediction of $1.26 \pm 0.19~\pb$.

\subsection{Electroweak tests}

Recent CDF and \DO measurements of diboson production cross sections and limits on trilinear gauge boson couplings using $1-5~\fbi$ have been presented~\cite{sekaric}. The diboson production is an important background for top quark, Higgs boson and SUSY particle studies. Precise knowledge and modeling of the diboson processes is vital for the current and future experiments. $Z \gamma$, $ZZ$, $WW$ and $WZ$ production cross sections have been measured and found to be in agreement with the SM next-to-leading order predictions. The \DO measurements were also used to set limits on anomalous triple gauge boson (TGC) couplings. The limits on the couplings, obtained with two different assumption on the relations between the anomalous TGCs, were the most stringent at a hadron collider, with sensitivity comparable to that of an individual LEP-II experiment. The \DO experiment also presented the world's best results on the $W$ boson magnetic dipole and electromagnetic quadrupole moments: $\mu_W = 2.02 ^{+0.08}_{-0.09} (e/2m_W)$ and $q_W = -1.00 \pm 0.09 (e/m_W^2)$.

As already mentioned, the measurement of $WW$ and $WZ$ production is relevant at the LHC because it is an irreducible background for Higgs and SUSY searches. The strategy for the measurement has been studied at CMS by means of Monte Carlo simulations~\cite{fabozzi}. $WW$ production has been studied by selecting events where both the $W$ bosons decayed leptonically, thus having two unlike-sign final state leptons and large MET. Also $WZ$ production has been studied, where the $W$ decays leptonically and the $Z$ decays into a lepton pair. The final state thus had three leptons and large MET. The cross section of $WW$ production could be measured with only $100~\pbi$ at $\sqrt{s} = 10~\Tev$, having a total uncertainty of 28\%. The $WZ$ signal could be observed with a statistical significance of $5 \sigma$, at 95\% C.L., with less than $350~\pbi$ at $\sqrt{s} = 14~\Tev$.

Events with at least two high-$p_T$ leptons (electrons or muons) have been studied by the H1 and ZEUS experiments at HERA with an integrated luminosity of $0.94~\fbi$~\cite{parenti}. The clean experimental signature, together with the precisely calculable small SM cross section, provides high sensitivity to possible contributions of physics beyond the SM. The total event yields were in agreement with the SM predictions, therefore the total and differential cross sections were measured and also found to be in agreement with the SM prediction. Seven events were observed in $e^+p$ collisions, having a scalar sum of all lepton's $p_T$, $\sum p_T$, greater than $100~\Gev$, while $1.94 \pm 0.17$ were expected; no such events were observed in $e^-p$ collisions, where $1.19 \pm 0.12$ were expected.

\subsection{Search for the Higgs boson}
\begin{figure}[!ht]
  \centering
  \includegraphics[width=0.30\textheight]{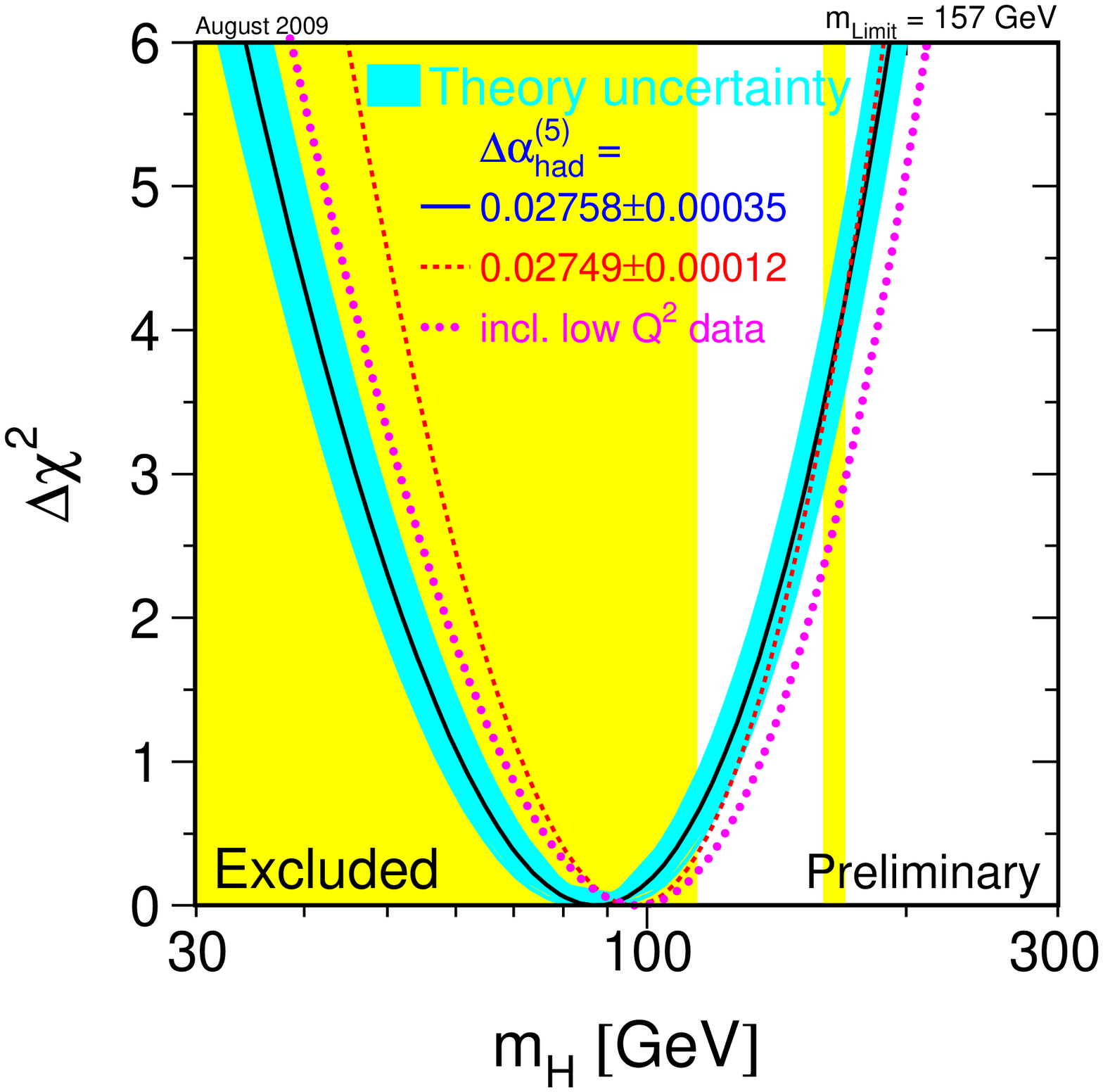}
  \caption{$\Delta \chi^2 = \chi^2 - \chi^2_{min}$ vs. $m_H$ curve. The line is the result of the fit using all high-$Q^2$ data (last column of Table 2 in~\cite{Alcaraz:2009jr}); the band represents an estimate of the theoretical error due to missing higher order corrections. The vertical band shows the 95\% C.L. exclusion limit on $m_H$ from the direct searches as of the summer 2009 combinations, at LEP-II (up to $114~\Gev$) and the Tevatron ($160~\Gev$ to $170~\Gev$). The dashed curve is the result obtained using a different evaluation of $\Delta \alpha^{(5)}_{had}(m^2_Z)$. The dotted curve corresponds to a fit including also the low-$Q^2$ data.}
  \label{fig:higgs:09}
\end{figure}
The low-mass region for the Standard Model Higgs boson is currently favoured by LEP, SLD and Tevatron experiments~\cite{Alcaraz:2009jr}. As of the summer 2009, direct measurements from LEP-II and Tevatron experiments had excluded SM Higgs masses below $114.4~\Gev$ and between 160 and $170~\Gev$, at 95\% C.L. (Fig. \ref{fig:higgs:09}). From fits to the electroweak data also an upper limit of $186~\Gev$ could be obtained.

\begin{figure}[!ht]
  \centering
  \includegraphics[width=0.44\textheight]{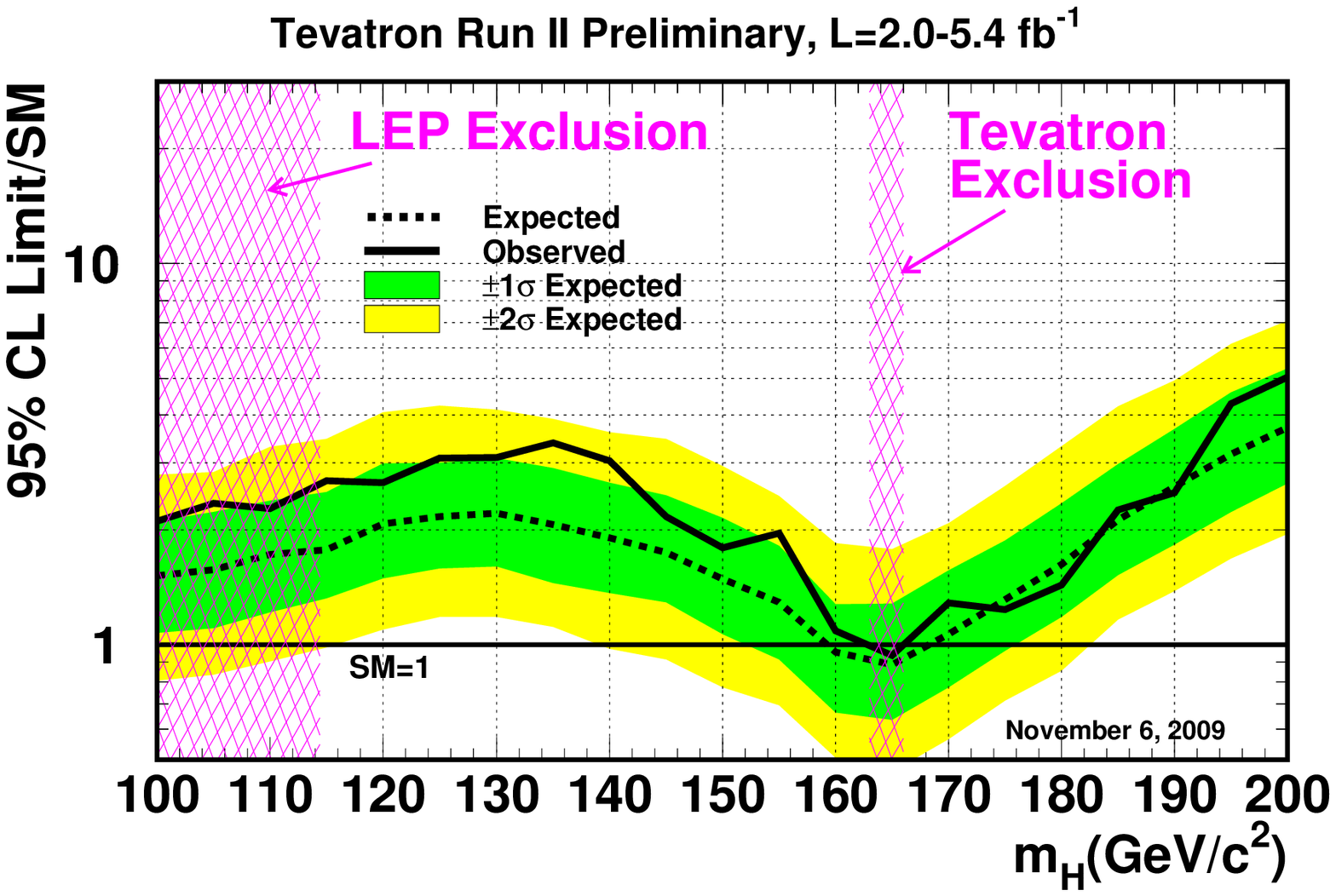}
  \caption{Observed and expected 95\% C.L. upper limits on the ratios to the SM cross section, as functions of the Higgs boson mass for the combined CDF and \DO analyses. The limits are expressed as a multiple of the SM prediction for test masses (every $5~\Gev$) for which both experiments have performed dedicated searches in different channels. The points are joined by straight lines for better readability. The bands indicate the 68\% and 95\% probability regions where the limits can fluctuate, in the absence of signal. The limits displayed in this figure are obtained with the Bayesian calculation.}
  \label{fig:smhiggs}
\end{figure}

The direct search for Higgs at the Tevatron~\cite{herner} has been done by combining 36 final states from CDF (using $2.0-4.8~\fbi$ of data) and 54 from \DO (using $2.1-5.4~\fbi$ of data), obtaining limits on the total production cross section $\sigma \times B(H \to X)$ in $p \bar{p}$ collisions at $\sqrt{s} = 1.96~\Tev$, for $m_H$ in the range $100 - 200~\Gev$. The results were presented in terms of the ratio of obtained limits to cross section in the SM, as a function of Higgs boson mass. A value of the combined limit ratio which is less than or equal to one indicates that that particular Higgs boson mass is excluded at the 95\% C.L. The combinations of results of each single experiment, as used in the Tevatron combination, yielded the following ratios of 95\% C.L. observed (expected) limits to the SM cross section: 3.10 (2.38) for CDF and 4.05 (2.80) for \DO at $m_H = 115~\Gev$, and 1.18 (1.19) for CDF and 1.53 (1.35) for \DO at $m_H = 165~\Gev$. The ratios of the 95\% C.L. expected and observed limit to the SM cross section are shown in Figure \ref{fig:smhiggs} for the combined CDF and \DO analyses. The region of Higgs boson masses excluded at the 95\% C.L. thus obtained was $163 < m_H < 166~\Gev$ (while the range expected to be excluded was $159 < m_H < 168~\Gev$).

Additional Higgs bosons are expected in many extensions to the Standard Model, for example the MSSM or the next-to-minimal Supersymmetry (NMSSM). In the MSSM, five Higgs bosons are predicted: $h^0$, $H^0$, $A^0$, and $H^\pm$. The \DO collaboration has searched for MSSM Higgs in the $\tau \tau$,  $\tau \tau b$, $bbb$ and $bbbb$ final states. The $\tau \tau$ data from \DO and CDF have been combined. All the results are generally in good agreement with the SM predictions~\cite{hays}. The NMSSM predicts the existence of other two Higgs bosons, $a^0$ and $h^0_S$. The Tevatron experiments also searched for the $a^0$; a small excess of $h^0$ decaying into an $a^0$ pair has been observed by \DO (4 events observed, $1.1 \pm 0.2$ expected), whereas no excess has been found by CDF.

\begin{figure}[!ht]
  \centering
  \includegraphics[width=0.42\textheight]{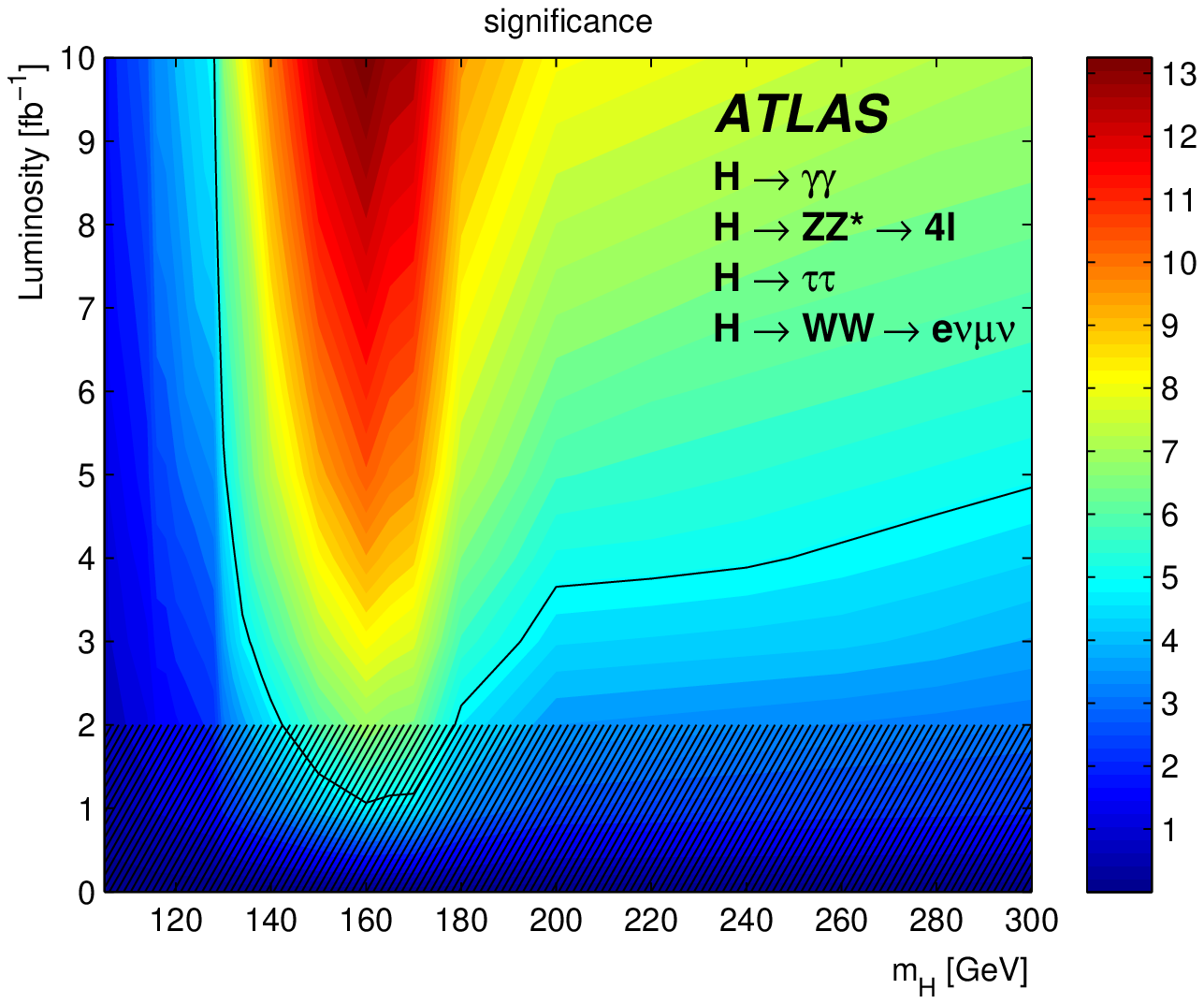}
  \caption{ATLAS combined discovery potential for Standard Model Higgs boson searches at $\sqrt{s} = 14~\Tev$.}
  \label{fig:higgs:atlas}
\end{figure}
It is well known that the search of the Higgs boson is one of the main goals of the LHC experiments. At ATLAS the various production and decay channels at $\sqrt{s} = 14~\Tev$ have been studied on Monte Carlo simulations~\cite{coniavitis}. By combining all the channels together the best sensitivity is achieved. As shown in Figure \ref{fig:higgs:atlas}, with $2~\fbi$ of data ATLAS has a $5 \sigma$ discovery sensitivity for a Standard Model Higgs in the mass range $143-179~\Gev$, while the exclusion sensitivity at 95\% C.L. reaches as low as $115~\Gev$. Also for MSSM Higgs bosons a substantial part of the SUSY parameter plane is expected to be covered.

\subsection{Top physics}

The first observation of single top production in $p \bar{p}$ collisions at the Tevatron has been reported~\cite{leone}.  Sophisticated multivariate techniques were needed in order to establish a tiny signal hidden by a large background. Both the CDF and \DO experiments have measured a single top cross section consistent with the SM prediction and with a significance larger than $5 \sigma$, and performed the first direct measurement of the CKM matrix element $| V_{tb} |$. The cross section measured by CDF (using $3.2~\fbi$ of data) and \DO (using $2.3-4.8~\fbi$ of data) have been combined, assuming a top mass of $170~\Gev$; the value obtained for the combined cross section is $\sigma = 2.76^{+0.58}_{-0.47}~\pb$.

At the LHC the cross section for single top production in $t$-channel is $124~\pb$, at $10~\Tev$ centre-of-mass energy; this is reduced by about a factor two at $7~\Tev$. At CMS single top candidates will be selected by asking a high-$p_T$ isolated muon in the barrel region, exactly two high-$p_T$ jets -- one of which must be tagged as $b$-jet -- and a large transverse mass of the $W$ boson candidate~\cite{wagner-kuhr}. Applying this selection, 102 single top and 229 background events (40\% of which are $t \bar{t}$ events) are expected with $200~\pbi$ at $\sqrt{s} = 10~\Tev$. The signal to background ratio is expected to remain about the same at $7~\Tev$.

The production of the top quark is also kinematically allowed at HERA, but the SM cross section is tiny (less than $1~\fb$). Flavour changing neutral current (FCNC) interactions could lead to a sizable single top production via an anomalous coupling, $\kappa_{tu \gamma}$, of the top quark to $u$-quarks and gluons. An upper limit at 95\% C.L. for the single top cross section was set by H1, $\sigma(ep \to etX) < 0.25~\pb$, and by ZEUS, $\sigma(ep \to etX) < 0.13~\pb$. The H1 cross section limit correspond to an upper limit on the coupling $\kappa_{tu \gamma} < 0.18$ (assuming $m_t = 175~\Gev$), whereas the ZEUS cross section limit corresponds to $\kappa_{tu \gamma} < 0.13~\pb$ (assuming a top mass of $171.2~\Gev$).

Many precision studies became possible at the Tevatron with the many thousand of top quarks collected~\cite{garberson}, for example the search for resonant $t\bar{t}$ production, the search for the top supersymmetric partner, $\tilde{t}$, and the search for a light charged Higgs in the top quark decays, $t \to H^+ b$. No anomalies have been found in such searches. The only slight deviation was found in the measurement of the top charge asymmetry: at the leading order the top quarks produced in the $p \bar{p} \to t \bar{t} X$ process will go equally often in the proton or in the anti-proton direction, but at the next-to-leading order an asymmetry $A_{fb} = 5 \pm 1\%$ is expected. The measurement from CDF, based on $3.2~\fbi$, is $A_{fb} = 0.193 \pm 0.065 (\mathrm{stat.}) \pm 0.024 (\mathrm{syst.})$, in agreement with the \DO result based on $0.9~\fbi$, $A_{fb} = 0.12 \pm 0.08 (\mathrm{stat.}) \pm 0.01(\mathrm{syst.})$.

At the Tevatron also many searches for anomalously produced top quarks have been performed~\cite{perfilov}. The search for an heavy gauge boson $W^\prime$ has been done, looking into the $W^\prime \to t b$ decay mode; no excess was found with respect to the SM expectations, therefore a lower limit on the mass of the resonance was derived. The CDF result is based on $1.9~\fbi$, that from \DO on $0.9~\fbi$; the most stringent limit at 95\% C.L. is from CDF, which excluded $M_{W^\prime} < 800~\Gev$ (if $M_{W^\prime} > M_{\nu_R}$), and $M_{W^\prime} < 825~\Gev$ (if $M_{W^\prime} < M_{\nu_R}$). Another important classes of processes which could contribute to top production are FCNC interactions of $u$- or $c$-quarks and gluons, where a top quark can be produced if the couplings $\kappa_{tug}$ or $\kappa_{tcg}$ are non-zero. The CDF results are based on $2.2~\fbi$, whereas those from \DO on $230~\pbi$ only. The exclusion limits at 95\% C.L. from CDF are $\kappa_{tug} < 0.018~\Tev^{-1}$, assuming $\kappa_{tcg}=0$, and $\kappa_{tcg} < 0.069~\Tev^{-1}$, assuming $\kappa_{tug}=0$. Other possible sources of top quarks that were investigated at the Tevatron are charged Higgs decays, $H^+ \to t \bar{b}$, and anomalous couplings of $W$, $t$ and $b$. Finally, the search for a fourth generation top quark was done by CDF based on $4.6~\fbi$ of data, and masses $m_{t^\prime}$ up to $335~\Gev$ were excluded at 95\% C.L.

\subsection{Searches}

One of the main goals of the LHC experiments is the search for physics beyond the Standard Model. The prospects for the discovery of R-parity conserving SUSY at ATLAS have been presented~\cite{borjanovic}. In R-parity conserving models, any supersymmetric particle would eventually decay into the lightest supersymmetric particle (LSP), after generating a cascade of ordinary particles; the LSP cannot decay any more, thus escapes detection. SUSY events will have MET and one or more jets (and possibly leptons) in the final state. At a centre-of-mass energy of $10~\Tev$ the discovery potential at $5 \sigma$ with $200~\pbi$ is up to masses of $600-700~\Gev$.

A search for R-parity violating SUSY has been performed with a data sample of $438~\pbi$ collected with the H1 detector at HERA~\cite{herbst}. This is the full data sample taken at $\sqrt{s} = 319~\Gev$. No evidence for squark production was found in the investigated final state topologies, therefore upper limits were set in the context of MSSM on the Yukawa couplings ($\lambda^\prime$) to ordinary particles. By assuming a Yukawa coupling of electromagnetic strength, $\lambda^\prime$ = 0.3, up-type squarks are excluded at 95\% C.L. up to $275~\Gev$, and down-type squarks up to $290~\Gev$.

The neutral current cross sections measured at HERA can be used to set limits on new interactions between electrons and quarks~\cite{panagoulias}. The investigated models were four-fermion contact interactions, leptoquarks, large extra dimensions, and quark substructure. Since no deviations were observed with respect to the SM predictions, limits were derived for all the models. For example, an upper limit on the effective quark-charge radius $R_q < 0.63 \cdot 10^{-16}$~cm was set by ZEUS, while H1 excluded $R_q < 0.74 \cdot 10^{-16}$~cm.

The most recent direct searches for new physics from BaBar and Belle were also presented~\cite{cervelli}. BaBar and Belle recorded respectively $486~\fbi$ and $798~\fbi$ around  the $\Upsilon(4\mathrm{S})$ resonance; moreover BaBar recorded $30~\fbi$ around the $\Upsilon(3\mathrm{S})$ resonance and $14~\fbi$ around the $\Upsilon(2\mathrm{S})$ in order to study $\Upsilon$ decays; Belle recorded also $121~\fbi$ around the $\Upsilon(5\mathrm{S})$ resonance, making it possible to study $B_s$ physics. Search for new physics was done in $B \to K^{(*)} \nu \bar{\nu}$ decays, $\tau$ decays, and  $\Upsilon(2\mathrm{S},3\mathrm{S})$ decays; no deviation from the SM were found, therefore upper limits on the branching ratios were set. BaBar has also searched for  next-to-minimal supersymmetric Higgs $A^0$ in the channels $\Upsilon(2\mathrm{S},3\mathrm{S}) \to \gamma A^0,~ A^0 \to \mu^+ \mu^-$, $\Upsilon(3\mathrm{S}) \to \gamma A^0,~ A^0 \to \tau^+ \tau^-$ and $\Upsilon(3\mathrm{S}) \to \gamma A^0,~ A^0 \to \mathrm {invisible}$, and set upper limits to the branching ratios.

The NA62 experiment at the CERN SPS has collected in 2007 the world largest sample of $K^\pm$ leptonic decays in order to test lepton universality, by measuring the ratio: $R_K = \Gamma(K^\pm \to e^\pm \nu)/ \Gamma(K^\pm \to \mu^\pm \nu) = \Gamma(Ke2)/\Gamma(K\mu 2)$~\cite{raggi}. The SM prediction for the ratio is very precise, thanks to the cancellation of hadronic effects: $R_K^{SM} = (2.477 \pm 0.001) \cdot 10^{-5}$. Effects from beyond the Standard Model could be sizable, for example in MSSM an enhancement of 1.3\% is expected. The precision of the world average is currently around 1\%, unable to exclude such an effect. The NA62 preliminary results, based on 51089 $K^\pm \to e^\pm \nu$ candidates, is $R_K = (2.500 \pm 0.016) \cdot 10^{-5}$, compatible with the Standard Model. The total uncertainty of the measurement is 0.6\%, still not enough to exclude MSSM effects. The aim is to reach a precision of 0.4\% with the full 2007-2008 statistics.

A large number of searches for physics beyond the Standard Model can be done with the first $10-100~\pbi$ of data collected by CMS at a centre-of-mass of $7~\Tev$~\cite{santanastasio}; the presented results were based on Monte Carlo events simulated with $\sqrt{s} = 10$ or $14~\Tev$, whose cross sections were rescaled to a $7~\Tev$ centre-of-mass energy. The sensitivity to contact interactions reachable by studying the ratio of di-jet events with both jets in the region $|\eta| < 0.7$ and di-jet events with both jets in $0.7 < |\eta| < 1.3$ is pretty high. Leptoquarks can be searched in the di-lepton plus di-jet channel; with $100~\pbi$ of collisions at $7~\Tev$ the sensitivity is already comparable to the one of the Tevatron with $1~\fbi$. Also the scan of the SUSY parameter space in the mSUGRA model with $100~\pbi$ of data shows that the sensitivity is comparable or better with respect to the Tevatron with $2-3~\fbi$ of data. An interesting search doable with the first data is that for heavy stable charged particles (HSCP); some models for new physics predicts the existence of heavy (hundreds of $\Gev$), long-lived (enough to decay outside the detector) charged particles. Heavy long-lived particles with hadronic nature, like gluinos or stops would hadronise in flight, forming meta-stable bound states with quarks and gluons (called R-Hadrons). Such particles could be distinguished from SM particles since they are slow (i.e. have small $\beta$) though having large momentum. The analysis results show that gluino and stop with masses of respectively about $500~\Gev$ and $350~\Gev$ could be excluded at 95\% C.L. with $100~\pbi$ of data at $\sqrt{s} = 7~\Tev$. R-Hadrons, due to their low $\beta$, could also be stopped in the CMS detector and decay later on, when there are no circulating beams; the signature would be an isolated jet in the hadronic calorimeter, in a detector that should be otherwise empty. A few weeks of data taking with an instantaneous luminosity of $10^{32} \mathrm{cm}^{-2} \mathrm{s}^{-1}$ would be sufficient to discover a long-lived gluino of $300~\Gev$ over a wide range of lifetimes.

\section*{Acknowledgements}
We would like to thank the organisers of the conference for their kind invitation, and also for making the conference a success despite the adversities. And we would also like to thank the speakers, some of which were travelling thousand of kilometres by train or car, or connecting in videoconference at pretty uncomfortable hours!

S.M. would like to thank A.P. for giving the summary of the theory talks on his
behalf during DIS 2010.

\bibliographystyle{parenti}
\bibliography{ew_searches_summary}

\end{document}